\documentclass[aps, twocolumn, prd, showpacs]{revtex4-1}
\usepackage{graphicx}
\usepackage{subfigure}
\usepackage{amsthm}
\usepackage{amssymb}
\usepackage{amsmath}
\usepackage{eucal}
\usepackage{natbib}
\usepackage{hyperref}
\newcommand{\dee}[2]{\frac{ \mathrm{d}#1 }{ \mathrm{d}#2 } }
\newcommand{\pd}[2]{\frac{\partial#1}{\partial#2}}

\begin{document}

\pagestyle{empty}
\title{Measuring emission coordinates in a pulsar-based relativistic positioning system}

\author{Darius Bunandar}
\email{darius.bunandar@utexas.edu}

\author{Scott A. Caveny}
\email{scaveny@physics.utexas.edu}

\author{Richard A. Matzner}
\email{matzner2@physics.utexas.edu}

\affiliation{Center for Relativity, The University of Texas at Austin, Austin, TX 78712-1081}

\date{\today}

\begin{abstract}
A relativistic deep space positioning system has been proposed using four or more pulsars with stable repetition rates.
(Each pulsar emits pulses at a fixed repetition period in its rest frame.)
The positioning system uses the fact that an event in spacetime can be fully described by emission coordinates: 
the proper emission time of each pulse measured at the event.
The proper emission time of each pulse from four different pulsars---interpolated as necessary---provides the four 
spacetime coordinates of the reception event in the emission coordinate system.
If more than four pulsars are available, the redundancy can improve the accuracy of the determination and/or resolve 
degeneracies resulting from special geometrical arrangements of the sources and the event.

We introduce a robust numerical approach to measure the emission coordinates of an event in any arbitrary spacetime geometry.
Our approach uses a continuous solution of the eikonal equation describing the backward null cone from the event.
The pulsar proper time at the instant the null cone intersects the pulsar world line is one of the four required coordinates.
The process is complete (modulo degeneracies) when four pulsar world lines have been crossed by the light cone.

The numerical method is applied in two different examples: measuring emission coordinates of an event in Minkowski 
spacetime using pulses from four pulsars stationary in the spacetime; and measuring emission coordinates of an event 
in Schwarzschild spacetime using pulses from four pulsars freely falling toward a static black hole.

These numerical simulations are merely exploratory, but with improved resolution and computational resources the 
method can be applied to more pertinent problems. For instance one could measure the emission coordinates, and 
therefore the trajectory, of the Earth.
\end{abstract}


\maketitle

\section{Introduction}
Pulsars---spinning neutron stars that emit directional electromagnetic radiation with an 
intriguingly stable period---in principle can be used as reliable interstellar lighthouses 
for spacecraft navigation in the Solar System and beyond 
%\cite{2002PhRvD..65d4017R, 2003jsrs...14...34C, coll}.
\cite{1981TDAPR..63...22C, 2005PhDT........30S, 2006JGCD, coll, ruggiero,2011AdSpR..47..645T, 2006ION}.
Pulsar spacecraft navigation has begun to be experimentally investigated by observing X-ray pulsars \cite{2006JGCD} 
from satellites, comparing these observations to satellite ephemerides.
Rovelli \cite{2002PhRvD..65d4017R}  suggested a method to construct this fully-relativistic 
universal coordinate system based on the proper emission times of the emissions from sources that 
he called ``satellites".
The concept of coordinates based on the emission times of pulses is identical in concept to the global positioning system (GPS). 
However, we treat a fully relativistic formulation, while the current GPS system treats relativistic effects only as 
perturbations from a Newtonian 
framework. A number of authors have developed the idea of a completely relativistic satellite positioning system; see 
\cite{2003jsrs...14...34C, 2009arXiv0905.3798T,2009arXiv0912.4418D,2010esaRpt}.
Delva \emph{et al.} \cite{2011arXiv1106.3168Dsummary} gave a recent review of this idea, including an extensive reference list.

Consider four pulsars, in motion in space, broadcasting pulses at a constant rate as measured in their proper times.
The intersections between the world lines of these pulsars and the past light cone of a reception event $\mathcal{R}$ give the proper 
emission times of the pulses from each pulsar that will be recorded at the event $\mathcal{R}$.
(We elaborate on the definition of the pulsar proper time, and on the interpolation between pulses in Section \ref{sec:gauge} below.)
The event $\mathcal{R}$, then, can be described by the coordinates $(\tau^1, \tau^2, \tau^3, \tau^4)$ called the 
\emph{emission coordinates}, where $\tau^1$ refers to the proper emission time of the pulse from pulsar \#1, $\tau^2$ 
refers to the proper emission time of the pulse from pulsar \#2, and so on.
A spacecraft recording the proper emission times of pulses from the four pulsars will then be able to determine its 
coordinates---and therefore its trajectory---in spacetime.

These emission coordinates can then in principle be converted into more conventional spacetime coordinates $(t, x, y, z)$. 
This paper provides a proof of concept numerical demonstration of determining the emission coordinates, and converting 
these coordinates to more standard spacetime coordinates.

Our method makes use of level set solutions of the eikonal equation describing the past light cone of the event $\mathcal{R}$. 
This method was originally developed by Caveny, Anderson, and Matzner to track black hole event horizons in computational 
simulation of black hole interactions \cite{cav}, and is similar to approaches in Refs. \cite{1994PhRvD..49.4004H} 
and \cite{1995PhRvL..74..630A}.
In the present paper, we show that the same numerical approach can address the problem of interconverting spacetime 
coordinates and the respective emission coordinates.
This approach is complete in the sense that a single method can be used to measure these emission coordinates even 
when the observer at $\mathcal{R}$---our spacecraft---and the pulsars are moving in an {\it arbitrary} spacetime geometry.

The problem of converting between conventional and emission coordinates naturally arises as one begins
to develop an intuition for emission coordinates. It was treated extensively by Delva and Olympio 
\cite{2009arXiv0912.4418D}; they have in mind that the source is a navigational satellite in the
Schwarzschild spacetime representing the Earth's gravitational field. 
Our eikonal solution for the backward null cone of the reception event $\mathcal{R}$ adds a 
new method to determine the emission coordinates of $\mathcal{R}$, in addition to those of \cite{2009arXiv0912.4418D}. 

Our method is given in a proof-of-principle form, with moderate computational accuracy; 
we discuss means to improve its accuracy (Section \ref{sec:compAccuracy} below). Its advantages are that it is 
general and robust in {\it any} 
spacetime (we give flat space and Schwarzschild examples);
it involves no ``shooting" or other iterative methods; it involves no approximations except 
discretization for computational integration; it builds up the entire past null cone; if the past null cone 
intersects the source world line, the intersection, and hence the emission coordinate, will be found.
To emphasize to the generality of the method: the Minkowski and the Schwarzschild examples differ
{\it only} in the metric used. The spinning Kerr spacetime could be treated similarly by inserting the Kerr
metric instead. And the method will straightforwardly work with a metric given only computationally, 
for instance the result of a simulation of the gravitational field of a binary pulsar.

This work is organized as follows:
Section \ref{sec:eikonal} outlines the theoretical framework.
Section \ref{sec:numerical} describes the implementation of the numerical description of the
eikonal equation. Our numerical approach uses a second-order artificial viscosity term, as in Caveny \emph{et al.} \cite{cav}.
Section \ref{sec:minkowski} presents the results of a numerical simulation where all pulsars are placed stationary in flat Minkowski space.
Section \ref{sec:schwarzschild} presents the results of a curved space numerical simulation where the pulsars are freely falling toward a Schwarzschild black hole. In both Section \ref{sec:minkowski} and Section \ref{sec:schwarzschild}, we actually measure the emission times from {\it five} pulsars. This allows us to construct a configuration which is 
easy to plot: four pulsars in the conventional coordinate plane $z=0$, a configuration which however has an obvious degeneracy between $\pm z$ when determining the emission coordinates (see Figures \ref{fig:minkowski} and \ref{fig:schwarzschild} below). The fifth pulsar is chosen out of the $z=0$ plane, and emission coordinates based on pulses from the fifth and any three of the first four pulsars are free of the $z=0$ degeneracy. {\it But} such a system is difficult to plot graphically, and we do not try. The multiple coordinate systems constructed in this way are related to one another by finite coordinate transformations.
Section \ref{sec:gauge} discusses this and the continuous {\it gauges} in which the emission coordinates are measured, and some improvements that are required when constructing a more practical pulsar-based positioning system.
Section \ref{sec:compAccuracy} discusses future improvements to the current code that will improve the code's numerical accuracy.
Conclusions are presented in Section \ref{sec:conclusion}.

We use geometrical units throughout, so the Newton constant $G$ and the speed of light $C$ are set equal to unity. We also use the Einstein summation convention so that repeated (one up, one down) indices are summed over their range. Greek lower case indices range and sum over $\{0,..,3\}$;
Latin lower-case indices range and sum over $\{1,2,3\}$.

%%%%%%%%%%%%%%%%%%%%%%%%%%%%%%%%%%%%%%%%%

\section{The eikonal equation}
\label{sec:eikonal}

See Ref. \cite{cav}.
The world line of a photon can be described by the equation of motion,
\begin{equation}
\dee{}{\tau} \left( \pd{L}{ \dot{x}^{\alpha}} \right) - \pd{L}{x^{\alpha}} = 0,
\end{equation}
where $\tau$ is an affine parameter and $\dot{x}^{\alpha} = \dee{x^{\alpha}}{\tau}$. Since the Lagrangian of a null geodesic motion, $L = \frac{1}{2} g_{\alpha \beta} \dot{x}^{\alpha} \dot{x}^{\beta} = 0$ has only kinetic terms, it is equal to the associated Hamiltonian---obtained using the Legendre transformation:
\begin{equation}
\label{eq:ham}
H = \frac{1}{2} g^{\alpha \beta} p_{\alpha} p_{\beta} = L.
\end{equation}

We introduce the 3+1 Arnowitt-Deser-Misner variables $\alpha$, $\beta_j$ and $\gamma_{ij}$,
%\footnote{We use Greek indices to denote spacetime components $0,1,2,3$ and Latin indices to denote spatial components $1,2,3$.}
%
\begin{equation}
\label{eq:ADM1}
\gamma_{ij} \equiv g_{ij}, \hspace{.5cm} \beta_{i} \equiv g_{ti}, \hspace{.5cm} \alpha^2 \equiv \beta_i \beta^i - g_{tt},
\end{equation}
where indices on $\beta_i$ are
raised by $\gamma^{ij}$ (the three-dimensional inverse of $\gamma_{ij}$) and lowered by $\gamma_{ij}$.
Equation (\ref{eq:ADM1}) implies
\begin{equation}
\label{eq:ADM2}
g^{tt} = - \frac{1}{\alpha^2}, \hspace{.5cm} g^{ti} = \frac{\beta^i}{\alpha^2}, \hspace{.5cm} g^{ij} = \gamma^{ij} - \frac{\beta^i \beta^j}{\alpha^2}.
\end{equation}

For a photon, which follows a null geodesic motion, the Hamiltonian has value zero and we can solve for $p_t$ to find
\begin{equation} \label{eq:mom}
p_t = \beta^i p_i \pm \alpha \sqrt{\gamma^{ij} p_i p_j}.
\end{equation}

The eikonal equation can be obtained by making simple direct substitutions
$p_t \rightarrow \pd{S}{t} = S_{,t}$ and $p_i \rightarrow \pd{S}{x^i} = S_{,i}$ in Eq. (\ref{eq:ham}) (dropping the factor of 1/2):
\begin{equation}
g^{\alpha \beta} S_{,\alpha} S_{,\beta} = 0
\end{equation}
which can be solved for $S_{,t}$. Using Eqs. (\ref{eq:ADM1}) and (\ref{eq:ADM2}), we obtain the following symmetric hyperbolic partial differential equation
\begin{equation} \label{eq:eikonal}
S_{,t} = \beta^i S_{,i} \pm \alpha \sqrt{\gamma^{ij} S_{,i} S_{,j}} = - \overline{H}.
\end{equation}
The bar is used to distinguish the Hamiltonian used here from the Hamiltonian in Eq. (\ref{eq:ham}). $\overline{H}$ is homogeneous of degree 1 in $S_{,i}$. The characteristic curves along which the level sets $\Gamma$ of $S$ are propagated are therefore
\begin{equation}
\dee{x^i}{t} = - \beta^i \mp \alpha \frac{ \gamma^{ij} p_j}{\sqrt{\gamma^{kl}p_k p_l}} = \pd{\overline{H}}{S_{,i}}
\end{equation}
and
\begin{equation}
\dee{S_{,i}}{t}= -\pd{}{x^i} \left( \beta^k S_{,k} \pm \alpha \sqrt{\gamma^{kj} S_{,k} S_{,j}} \right) = - \pd{\overline{H}}{x^{i}},
\end{equation}
which are the null geodesic equations. The integral curves of the gradients of $S$ and $\Gamma$ are then also the null geodesics:
\begin{equation}
\dee{x^i(\lambda)}{\lambda} = g^{i \alpha} p_{\alpha} = g^{i \alpha} S_{,\alpha} = S^{,i}(\lambda, x^j (\lambda))
\end{equation}
where $\lambda$ is the affine parameter in this case.

Solutions to the eikonal equation fall into topologically equivalent classes. This means that any smooth function $\psi(S)$ topologically equivalent to $S$ is also a solution. Note the following relations,
\begin{equation} \label{eq:reinit1}
\psi_{,i} (S) = \pd{\psi}{S} \pd{S}{x^i} = \lambda(S) S_{,i}
\end{equation}
and
\begin{equation} \label{eq:reinit2}
\psi_{,t} (S) = \pd{\psi}{S} \pd{S}{t} = \lambda(S) S_{,t}.
\end{equation}

The above relations are true because Eq. (\ref{eq:eikonal}) is homogeneous of degree 1 in momentum. The above results guarantee that a smoothly related initial data $S_0 \rightarrow S_0' = \psi(S_0)$ have smoothly related solutions.

The solutions of the eikonal equation also guarantee the equivalence of ingoing and outgoing solutions under time reversal. Referring to Eq. (\ref{eq:eikonal}), propagation of data for $S$ describing an ingoing or outgoing null surface is completely specified by:
\begin{enumerate}
\item a definition of the direction of time,
\item choices of $\alpha$ and $\beta^i$, and
\item a choice of the sign of the root.
\end{enumerate}
%%%%%%%%%%%%%%%%%%%%%%%%%%%

\section{Numerical method using the eikonal}
\label{sec:numerical}
Our numerical method makes use of the time evolution equation for the solutions $S$,
\begin{equation} \label{eq:evolve}
S_{,t} = - \overline{H}(t,x^i, S_{,j}),
\end{equation}
which allows for relatively fast calculations while being sufficiently accurate.

The eikonal equation, however, shows singular behavior and as described by Ehlers and Newman \cite{newman}, the eikonal equation generally breaks down on caustic and other sets. To address this problem, we make use of an explicit artificial viscosity term to control the appearances of such singularities. Adding the artificial viscosity at the level of the finite difference approximation corresponds to replacing the time evolution Eq. (\ref{eq:evolve}) with the evolution equation
\begin{equation} \label{eq:visc}
S_{,t} = \epsilon \nabla^2 S - \overline{H}(t,x^i, S_{,j})
\end{equation}
where $\epsilon$---the artificial viscosity---is a small quantity of the order of $h^2$ ($h$ denotes the resolution of the numerical mesh) and $\nabla^2$ is any second-order, linear derivative operator. We use a second-order finite difference approximation to the Laplacian.

The null surface $\Gamma$, at any given time level, can be extracted from the level set section of the eikonal solution $S$, say $S = 1$. This problem of extraction is an inverse problem, since it requires that points $(x, y, z)$ are found such that $S(x,y,z) = 1$. Nevertheless, a combination of ordinary bisection and interpolation method is sufficient to extract the approximate null surface $\hat{\Gamma}$. In this method, the surface $\hat{\Gamma}$ can be represented in spherical coordinates $( u(\theta, \phi), \theta, \phi)$, where $r = u(\theta, \phi)$ is the surface function for a given center $c^i$ contained within the surface $\hat{\Gamma}$.
%%%%%%%%%%%%%%%%%%%%%%%%%

\section{Numerical simulation in Minkowski space}
\label{sec:minkowski}
The first application of our numerical method to measure emission coordinates is done in the simplest configuration possible: the pulsars are stationary in flat Minkowski space and we are measuring the emission coordinates of an event $\mathcal{R}$ at the spatial origin $(t, x, y, z) = (0.1, 0, 0, 0)$. Although Minkowski space has no natural timescale, we may take the coordinates to have units of seconds.

As described in Section II, defining the numerical simulation requires the choices of the direction in time, $\alpha$, $\beta^i$ (determined from our Minkowski coordinates as $\alpha = 1$ and $\beta_i=0$), and the sign of the root in Eq. (\ref{eq:eikonal}). To measure the emission coordinates---the intersections between the past light cone and the world lines of the pulsars---we need to choose the sign of the root to be negative so that the
propagation of $S$ describes an outgoing null surface when the direction of time is pointing to the past.

The simulation is done in a three-dimensional computational domain of $N^3$ points with $N = 361$. The outer boundaries are located at $[ -2.5, +2.5 ]$ in the $x, y, z$ directions.
The resolution of this finite difference mesh is then $h = 5/360$ sec.
A Courant-Friedrichs-Lewy factor of $\lambda = 1/4$ with an iterated Crank Nicholson scheme \cite{teu} is used in the finite difference approximation of the time evolution equation (\ref{eq:visc}). Here the artificial viscosity parameter $\epsilon$ is set to be equal to $h^2/16$. We found that applying $\approx$ 400 numerical evolution steps (stopping at the time slice $t = -1.388$ sec) was more than enough to measure the emission coordinates.

\begin{figure}[hbtp!]
 \centering
 \includegraphics[width=.48\textwidth]{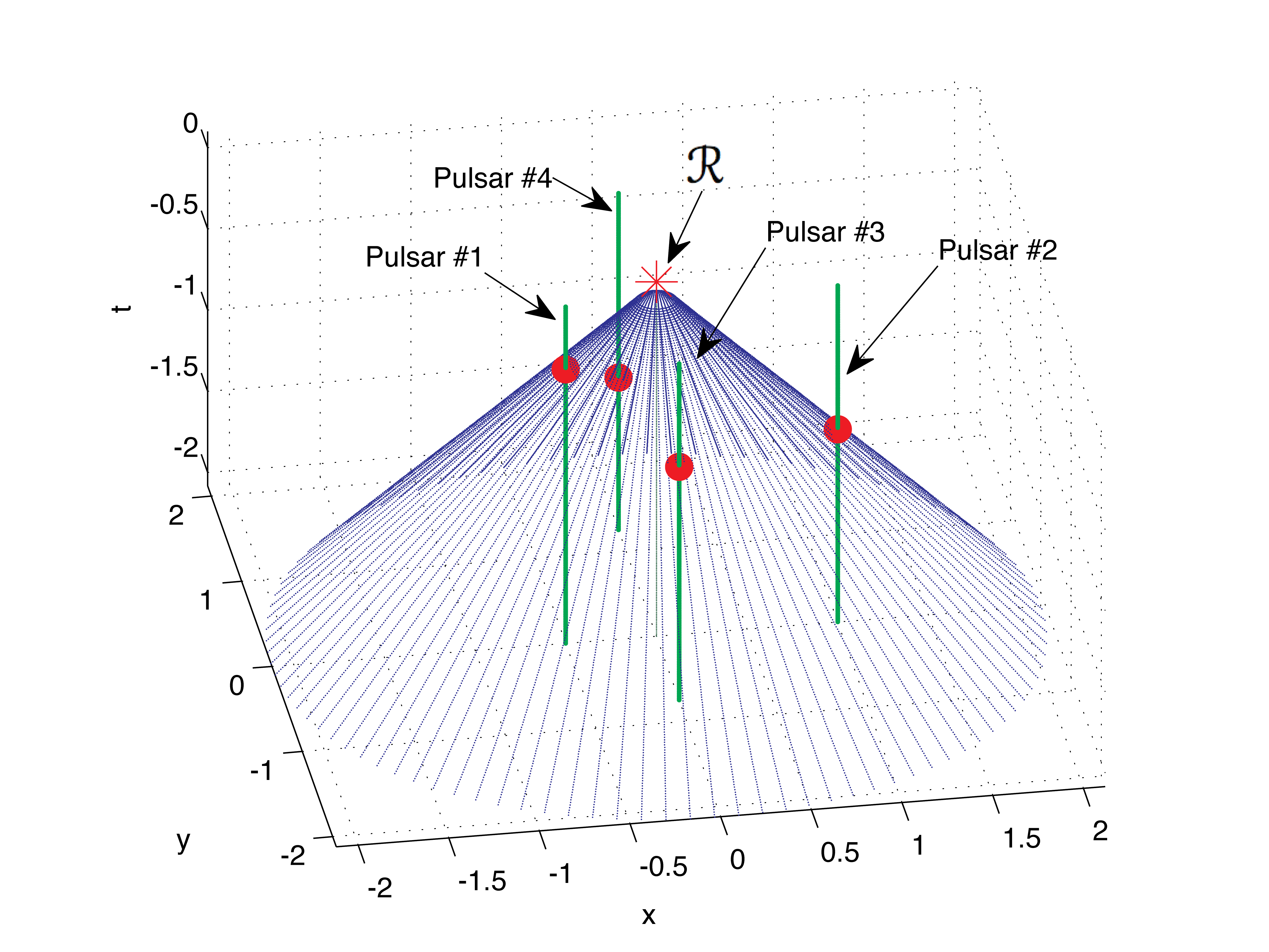} % requires the graphicx package
 \caption{Rays: plot of null surface $\hat{\Gamma}$ (the past light cone of $\mathcal{R}$
in Minkowski spacetime).
Star marks the location of the event $\mathcal{R} = (t,x,y,z) = (0.1, 0, 0, 0)$ that we are coordinatizing.
Bold lines: world lines of the four stationary $z=0$ pulsars in Minkowski spacetime. 
Dots mark the intersections between the null surface and the pulsars.
The coordinates of the intersections are recorded in Table \ref{table:minkowski}.
Analytic methods are used to substitute for the null cone between $\mathcal{R}$ and the top of the computed (truncated) null cone. See the discussion in the text.}
 \label{fig:minkowski}
\end{figure}

In this simulation, the stationary pulsars are located at:
\begin{enumerate}
\item pulsar \#1: $(t, x, y, z) = (t, -0.50, 0, 0)$,
\item pulsar \#2: $(t, x, y, z) = (t, 1.00, 0, 0)$,
\item pulsar \#3: $(t, x, y, z) = (t, 0, -0.75, 0)$,
\item pulsar \#4: $(t, x, y, z) = (t, 0, 1.25, 0)$, and
\item pulsar \#5: $(t, x, y, z) = (t, 0.30, 0.40, 0.50)$.
\end{enumerate}
The first four create a quadrilateral in the $z=0$ plane surrounding the observer at $x = y = z = 0$, which will generally produce a $\pm z$ degeneracy when determining the emission coordinates.
This configuration is easy to plot; see Figure \ref{fig:minkowski}.
Including pulses from the fifth pulsar with those from three of the others create emission coordinates without the $\pm z$ degeneracy. Alternately one can combine data from more than four (e.g. five) pulsars by a kind of least squares fitting. That is the subject of work by Tarantola \emph{et al.} \cite{2009arXiv0905.3798T}. We assume that the proper time of each pulsar is precisely the Minkowski coordinate time $t$; this choice is available only in Minkowski spacetime, but not in curved spacetime (see Sections \ref{sec:schwarzschild} and \ref{sec:gauge}).

In order to avoid the focal point singularity at the vertex $\mathcal{R}$ of the null cone,
we represent the event by positing a spherically symmetric null surface $\Gamma$ centered at the origin with radius $\rho = 0.1$ sec at time $t=0$ sec. We thus assume (trivially correct in Minkowski space) that near the event under consideration, spacetime is sufficiently flat so that the light cone is spherical around the event, and set the data a small amount of time (here $0.1$ sec) in the past of the event we are coordinatizing.

Data for the eikonal equation are set in the spherically symmetric form
\begin{equation} \label{eq:initialdata}
S(t = 0, x^i) = 1 + \tanh{\left( \frac{r_c - r}{c} \right)}
\end{equation}
where $r_c$ is the radius of the initial surface $\Gamma$ and is equal to $0.1$ sec in this case, 
while $c$ controls the steepness of the hyperbolic tangent function. We set $c$ to $0.1$ sec in our experiment.

After every 75 iterations, we reinitialize the data for the eikonal equation with a function that is similar to Eq. (\ref{eq:initialdata}),
\begin{equation}
\label{eq:numReinit}
S(t, x^i) = 1 + \tanh{ \left( \frac{u(\theta, \phi, t) - r}{c'} \right)}
\end{equation}
to ensure the smoothness of the data: reinitialization allows us to set viscosity parameter $\epsilon$ to be 
arbitrarily small while avoiding the onset of singularity.
Here $c'$ denotes a new steepness of the function and our simulation uses $c' = c$.  Recall that 
$u(\theta, \phi, t)$ is the surface function for a given center $c^i$ contained within the discrete null surface $\hat{\Gamma}$.

We know for a fact that the reinitialized solution $S$ is also a solution to the eikonal equation by 
virtue of Eqs. (\ref{eq:reinit1}) and (\ref{eq:reinit2}). What Eq.(\ref{eq:numReinit}) accomplishes 
is to smooth the function $S$ near the location of the null surface (where $S=1$) to produce 
data for a solution which are both analytic and smooth, and which describe the same null cone. 

For the resolution used in our simulations, we find no difference in behavior in the Minkowski case, 
whether or not this reinitialization is carried out. However, in the Schwarzschild curved spacetime case, the 
reinitialization is necessary, as discussed below.

Table \ref{table:minkowski} shows the results of our measurements
to determine the emission coordinates of the event $(t, x, y, z) = (0.1, 0, 0, 0)$.

Figure \ref{fig:minkowski} depicts the intersections of the past light
cone generated by our numerical simulation in the equator $(z = 0)$ with
the world lines of pulsars \#1 through \#4, depicted by the four bold lines.
The dots mark the events when each pulsar emits its pulse, which will later be recorded by 
the observer at the event $\mathcal{R}$ we are coordinatizing.

\begin{table}[hbtp!]
       \centering
       \begin{tabular}{c c | c c c c}
       Pulsar \#       &       $\tau$  &       $t$     &       $x$     &       $y$     &       $z$ \\
       \hline \hline
       1 & $-0.399 \pm 0.003$ & $-0.399 \pm 0.003$ & -0.500 & 0.000 & 0.000 \\
       2 & $-0.903 \pm 0.003$ & $-0.903 \pm 0.003$ & 1.000 & 0.000 & 0.000 \\
       3 & $-0.653 \pm 0.003$ & $-0.653 \pm 0.003$ & 0.000 & -0.750 & 0.000 \\
       4 & $-1.156 \pm 0.003$ & $-1.156 \pm 0.003$ & 0.000 & 1.250 & 0.000 \\
       5 & $-0.615 \pm 0.003$ & $-0.615 \pm 0.003$ & 0.300 & 0.400 & 0.500 \\
       \end{tabular}
       \caption{Results of emission coordinates in Minkowski spacetime of an event $\mathcal{R}=(t, x, y, z) = (0.1, 0, 0, 0)$. $\tau$ denotes the proper time of the
pulsar when the pulsar world line intersects the observer's past light cone.
Any four of the five proper times listed in the table constitute the emission coordinates in 
a particular emission coordinate system, of the event point $\mathcal{R}$.
The coordinates $t, x, y$, and $z$ are the Minkowski space coordinates of the pulsars when the intersections occur.}
       \label{table:minkowski}
\end{table}

Comparing our results with analytical calculations, the errors in our
results are mainly of the order of a few time resolutions of the finite
difference mesh, $dt = \lambda \cdot h = 0.0035$. Simple analytical
calculations show that world lines of pulsars \#1, \#2, \#3,  \#4 and \#5 are
expected to intersect the null surface at $t = -0.400$ sec$, -0.900$ sec$, -0.650$ sec, $-1.150$ sec, and $-0.607$ sec,
respectively. The errors can be attributed to the numerical
accuracy of the simulations, mainly due to: (1) the interpolation routine
that extracts the discrete surface $\hat{\Gamma}$, and (2) the resolution
of the underlying three dimensional grid.
%%%%%%%%%%%%%%%%%%%%%%%%%%

\section{Numerical simulation in Schwarzschild geometry}
\label{sec:schwarzschild}
Our numerical method has been successful in measuring the emission coordinates
in Minkowski spacetime but in general spacetime is not flat.
To investigate the change in the emission coordinates in a curved spacetime, we
evaluate the numerical method in a Schwarzschild geometry containing
a stationary black hole of mass $M=0.25$.

We use the Eddington-Finkelstein \cite{1924Natur.113..192E, 1958PhRv..110..965F} 
coordinate system in describing the Schwarzschild geometry to avoid the coordinate singularity at areal distance $r = 2M$ from the black hole.
The standard Eddington-Finkelstein form of the Schwarzschild solution is centered at the spatial origin of the Eddington-Finkelstein coordinates.
In order to maintain our event position at $(x,y,z) = (0,0,0)$, we offset the black hole coordinates, putting it at $(x_0, y_0, z_0) = (2.5, 0, 0)$ to provide a strong gravitational attraction at $(0,0,0)$.
The Schwarzschild metric in Eddington-Finkelstein coordinates (in Kerr-Schild form) is \cite{2002PhRvD..66h4024H}:
\begin{equation}
g_{\alpha \beta} = \eta_{\alpha \beta} + \frac{2M}{r} l_\alpha l_\beta
\label{31}
\end{equation}
where  $r=\sqrt{(x-x_0)^2 +(y-y_0)^2 + (z-z_0)^2}$ is the areal coordinate distance from the center of the black hole.
Here $\eta_{\alpha \beta} = \text{diag}(-1, 1, 1, 1)$ is the Minkowski metric and $l_\alpha$ is an outgoing null vector with respect to both the Minkowski and the Schwarzschild metric;  the outgoing null vector written explicitly in terms of the coordinates $(t, x, y, z)$ has the form
\begin{equation}
l_\alpha \rightarrow \left( 1, \frac{x-x_0}{r},\frac{y-y_0}{r}, \frac{z-z_0}{r} \right).
\label{32}
\end{equation}

In Eddington-Finkelstein coordinates, the quantities $\alpha (> 0)$ and $\beta_i$ are:
\begin{equation}
\alpha^2 = \frac{1}{1+2M/r},
\label{33}
\end{equation}
and
\begin{equation}
\beta_i = \frac{2M}{r^2}\left(x-x_0, y-y_0 , z-z_0\right).
\label{34}
\end{equation}

The computational parameters of the simulation are the same as in the previous section:
$N^3$ points with $N = 361$; outer boundary locations at $[-2.5, +2.5]$
in the $x, y, z$ directions; a Courant-Friedrichs-Lewy factor
of $\lambda = 1/4$; and artificial viscosity parameter of $\epsilon = h^2/16$.
However, $\approx 550$ evolution steps to the $t = -1.910$ sec time slice are needed in 
this particular case because of the shear of the null cone and the movements of the pulsars,
 both of which will be discussed later in this section.

As noted in Section \ref{sec:minkowski}, the vertex of the null cone is difficult to handle 
because of the finite resolution of the grid used to evolve the null cone into the past.
In the flat Minkowski space treated in Section \ref{sec:minkowski}, the null cone can be 
described analytically, and is a shear-free metric sphere expanding from the vertex, with 
radius equal to the elapsed Minkowski time.
Thus in Section \ref{sec:minkowski} we set the data at a Minkowski time that is $0.1$ sec 
earlier ($\Delta t = -0.1$ sec) than the event we are coordinatizing---when the backward 
light cone sphere had a radius of $0.1$ sec.
This sphere is sufficiently large so that our discretization adequately resolves it, and 
as we follow the expanding light cone into the past, the relative resolution becomes even 
better. In the Minkowski case this data setting method contributes to minimal total errors 
(a few times the discretization size).

We use a similar method to initialize the null ``cone" here in the black hole spacetime.
(The null cone here is not a spherical cone because of shear due to the presence of 
the black hole.) Again we analytically set the boundary condition at a small coordinate 
time ($0.1$ sec) in the past of the event we are coordinatizing.
To provide a technique for general spacetime, we define an approximate method that will 
work in any spacetime:
we evaluate the metric at the event being coordinatized, and assume that it is constant in 
the small region needed to propagate the null surface backward for a small arbitrary coordinate time.

In the current example, we will find the emission coordinates of the event given 
by $(t, x, y, z)=(0.1,0,0,0)$ in the Eddington-Finkelstein coordinates where the 
black hole is offset to $(x_0, y_0, z_0) = (2.5,0,0)$ and has a mass $M = 0.25$.
Therefore, $2M/r= 1/5$ at $(0.1,0,0,0)$, and from Eqs. (\ref{31}), (\ref{32}) the metric at that point is
$g_{tt}=-4/5;$ $g_{tx}=-1/5;$ $g_{xx}=6/5;$ $g_{yy}=g_{zz}=1$.

One proceeds by choosing a small interval of Eddington-Finkelstein coordinate time 
$\Delta t$ (here $-0.1$ sec) and then solving for the values of the 3-space points ($\Delta x,\Delta y, \Delta z$) that satisfy
\begin{equation}
0= g_{tt}\Delta t^2 +2g_{tx}\Delta t \Delta x + g_{ij} \Delta x^i \Delta x^j,
\end{equation}
using the constant values of the metric coefficients defined above.
The 3-space points found define the shape of the $t=$ constant $= \Delta t$ slice of the backward null 
cone, which initializes the eikonal data for further evolution into the past.

%Also, the value of the proper time elapsed $\Delta \tau$ is now different for each pulsar 
%because %its value is dependent on the distance of the pulsar from the black hole.
%Nevertheless, we can calculate its value for the approximate coordinate time 
%elapsed using the 
%parametric relationships between the two that results from considering the geodesic motion of the 
%pulsars (Eqs. (\ref{eq:taueta}), (\ref{eq:tbareta}) below).

Contrary to the Minkowski case, now the pulsars are freely falling toward the black hole located at $(x_0, y_0, z_0) = (2.5, 0, 0)$.
As an initial condition we demand that at time $t = 0$ sec, all five pulsars are located at coordinate positions with the same values as used in the previous section, with zero velocity.
However, due to black hole gravitational acceleration, the pulsars are moving at other times.
Because they begin at rest in the black hole frame, the pulsars have only a radial velocity---and no angular velocity---toward the black hole.
Therefore, to measure the emission coordinates of an event at the origin, we need to first understand the radial geodesic motion of the freely-falling pulsars.

The Schwarzschild metric written in spherical Eddington-Finkelstein coordinates is
\begin{align} \label{eq:ef}
\mathrm{d}s^2 = &- \left( 1- \frac{2M}{r} \right) \mathrm{d}t^2 + \frac{4M}{r} \; \mathrm{d}r\mathrm{d}t + \left( 1 + \frac{2M}{r} \right) \mathrm{d}r^2  \nonumber \\
&+ r^2 ( \mathrm{d}\theta^2 + \sin^2{\theta} \; \mathrm{d}\varphi^2 )
\end{align}
and the Schwarzschild metric written in the Schwarzschild coordinates is
\begin{align} \label{eq:schwarzschild}
\mathrm{d}s^2 = &- \left( 1- \frac{2M}{r} \right) \mathrm{d}\bar{t}^2 + \left( 1- \frac{2M}{r} \right)^{-1} \mathrm{d}r^2 \nonumber \\
&+ r^2 ( \mathrm{d}\theta^2 + \sin^2{\theta} \; \mathrm{d}\varphi^2 ).
\end{align}
The bar on $\bar{t}$ here is used to distinguish the Schwarzschild coordinate time $\bar{t}$ from the Eddington-Finkelstein coordinate time $t$.

From the expressions of Schwarzschild metric in the two coordinate systems, it is clear that the area coordinate $r$ in Eddington-Finkelstein coordinates is the same area coordinate $r$ in Schwarzschild coordinates.
This guarantees that a pulsar's radial geodesic motion found using Schwarzschild coordinates, when expressed in terms of the pulsar proper time $\tau$, will have identically the same expression in Eddington-Finkelstein coordinates.
Let us proceed in the Schwarzschild coordinates.

Purely radial geodesic motion of a freely falling pulsar can be described by the parametric equations \cite{mtw}:
\begin{equation} \label{eq:reta}
r = \frac{R}{2}( 1 + \cos{\eta} ),
\end{equation}
\begin{equation} \label{eq:taueta}
\tau = \frac{R}{2} \left( \frac{R}{2M} \right)^{1/2} ( \eta + \sin{\eta} ),
\end{equation}
and
\begin{widetext}
\begin{equation} \label{eq:tbareta}
\bar{t} = \left[ \left( \frac{R}{2} + 2M\right) \left( \frac{R}{2M} - 1 \right)^{1/2} \right] \eta  + \frac{R}{2} \left( \frac{R}{2M} - 1 \right)^{1/2} \sin{\eta} + 2M \ln \left| \frac{ (R/2M -1)^{1/2} + \tan{(\eta/2)} }{ (R/2M -1)^{1/2} - \tan{(\eta/2)} }\right|
\end{equation}
\end{widetext}
where $\eta$ is the parameterization, and $R$ is the apastron---the areal distance at which the pulsar has zero velocity. Note that both $\bar t=0$ and $\tau =0$ when $\eta =0$. As a specific choice in our simulation, we also specify that $\tau=0$, for each pulsar, occurs at Eddington-Finkelstein coordinate time $t=0$.

We have obtained the expression for Schwarzschild time, but the Schwarzschild metric we supplied to the numerical simulation is expressed in Eddington-Finkelstein coordinates.
Thus, we need to relate the Eddington-Finkelstein time and the Schwarzschild time, which can be done by relating the spacetime interval in Eqs. (\ref{eq:ef}) and (\ref{eq:schwarzschild}) to obtain
\begin{equation}
\mathrm{d}\bar{t}^2 = \left[ \mathrm{d}t - \frac{2M/r}{1-2M/r} \; \mathrm{d}r \right]^2.
\end{equation}
Taking the positive root of this equation and integrating, we obtain the expected expression:
\begin{equation}
t = \bar{t} + 2M \ln{(r - 2M)} + C
\end{equation}
where $C$ is an arbitrary constant that depends on the initial condition.
In our simulation, the initial condition is at $t = \bar{t} = 0$, $r = R$.
Substituting this and solving for $C$, we obtain the final equation that relates the Schwarzschild time $\bar{t}$ with the Eddington-Finkelstein time $t$,
\begin{equation} \label{eq:teta}
t = \bar{t} + 2M \ln{(r - 2M)} - 2M \ln{(R - 2M)}.
\end{equation}
Notice that $R$ depends on which pulsar is under consideration, so we are setting data that have $t=0=$ constant, but have a different $\bar{t}$ for each pulsar. We now have the complete relations needed to describe the purely radial geodesic motion of the pulsars.

\begin{figure}[hbtp!]
 \centering
 \includegraphics[width=.48\textwidth]{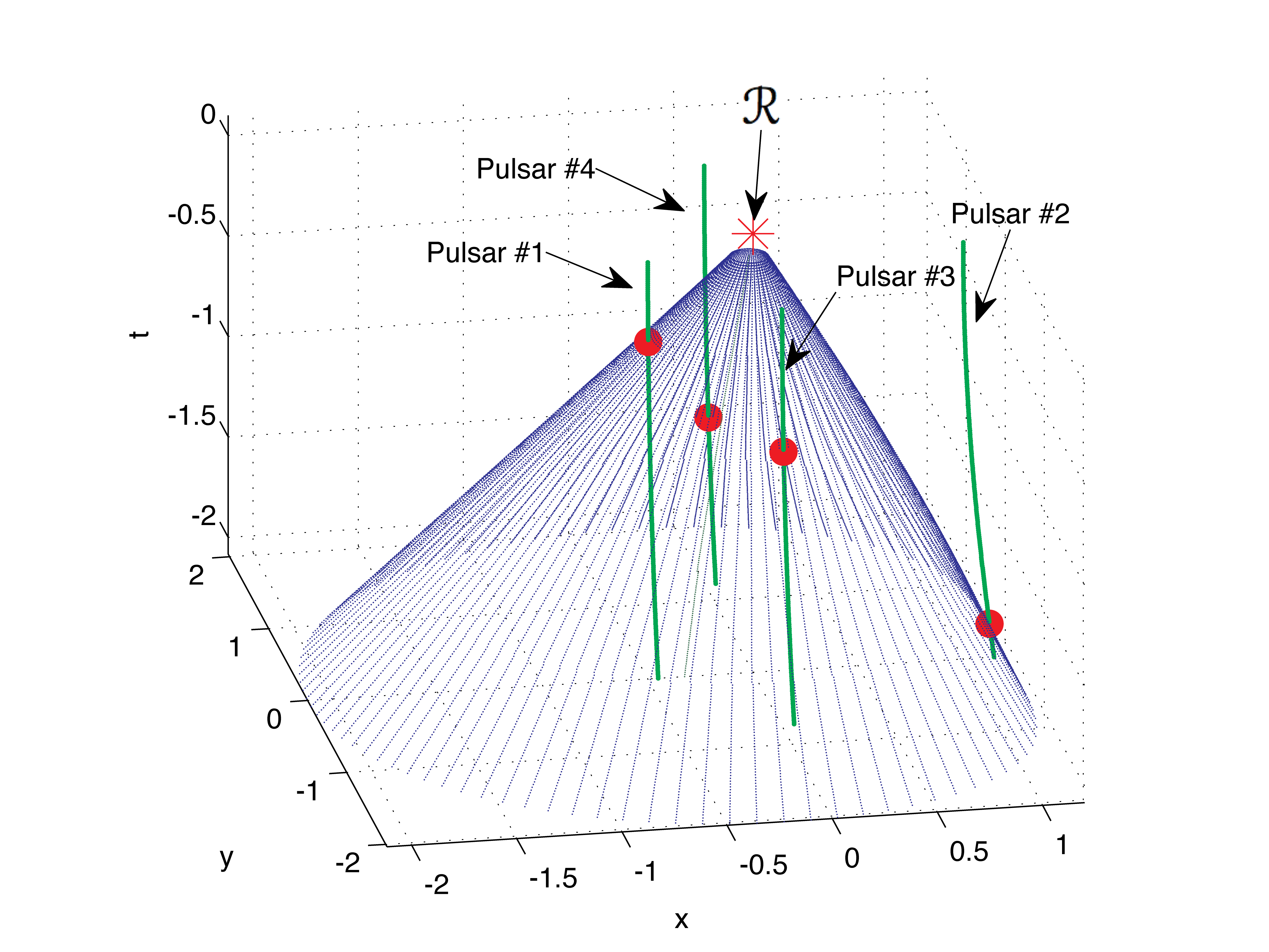} % requires the graphicx package
 \caption{Rays: the past null cone  $\hat{\Gamma}$ of the measurement event $\mathcal{R}$. We use Eddington-Finkelstein coordinates in a Schwarzschild spacetime.
Star marks the location of event $\mathcal{R} = (t,x,y,z) = (0.1, 0, 0, 0)$ that we are coordinatizing.
Bold lines: world lines of four $z=0$ pulsars freely falling toward a stationary black hole located to the right of the volume plotted at $(x_0, y_0, z_0) = (2.5, 0, 0)$.
Dots mark the intersections between the null surface and the pulsar world lines.
The coordinates of these intersections are recorded in Table \ref{table:schwarzschild}.
See the text for a discussion of the gap between $\mathcal{R}$ and the top of the (truncated) null cone.}
\label{fig:schwarzschild}
\end{figure}

As we run our simulation to obtain the null surface $\hat{\Gamma}$, the pulsars move along the geodesics as described by Eqs. (\ref{eq:reta}), (\ref{eq:tbareta}), and (\ref{eq:teta}).
Figure \ref{fig:schwarzschild} depicts the intersections of the past light cone generated by 
the simulation at the equator, $z = 0$.
The world lines of the first four pulsars are again depicted by four bold lines and their 
intersections with the null cone by dots.

The curvature of pulsar \#2's world line is most evident in Figure \ref{fig:schwarzschild}. Note also in Figure \ref{fig:schwarzschild}, the shape of the light cone: the light cone moves further away from the black hole as we go further into the past, as expected from intuition. (It is falling into the black hole as time goes forward.) The cone also flattens out as we go further into the past due to the tidal field of the black hole. These movements of the light cone are a main reason why the measured emission coordinates in Table \ref{table:schwarzschild}, the results of the measurements in Schwarzschild geometry, differ from the results in Table \ref{table:minkowski}.
The proper times $\tau^\alpha$ in Table \ref{table:schwarzschild} could in principle be obtained by integration back along the pulsar world lines. However, with the analytic results, $\tau^\alpha$ are in fact obtained by computing $\eta$ for each pulsar at the event at which it intersects the null cone from Eqs. (\ref{eq:reta}), (\ref{eq:tbareta}), and (\ref{eq:teta}), and then obtaining $\tau$ from Eq. (\ref{eq:taueta}).

Although we do not develop a complete analytical solution for the photons in this case
(see \cite{2009arXiv0912.4418D} for such a solution), the configuration is qualitatively comparable to the Minkowski case discussed earlier. Therefore, we assume a similar level of error in the Schwarzschild case and indicate that in Table \ref{table:schwarzschild}.

\begin{table}[hbtp!]
       \centering
       \begin{tabular}{c c | c c c c}
       Pulsar \#       &       $\tau$  &       $t$     &       $x$     &       $y$     &       $z$ \\
       \hline \hline
       1 & $-0.364 \pm 0.003$ & $-0.399 \pm 0.003$ & -0.498 & 0.000 & 0.000 \\
       2 & $-1.432 \pm 0.003$ & $-1.840 \pm 0.003$ & 1.170 & 0.000 & 0.000 \\
       3 & $-0.641 \pm 0.003$ & $-0.715 \pm 0.003$ & 0.007 & -0.748 & 0.000 \\
       4 & $-1.112 \pm 0.003$ & $-1.233 \pm 0.003$ & 0.017 & 1.241 & 0.000 \\
       5 & $-0.618 \pm 0.003$ & $-0.701 \pm 0.003$ & 0.309 & 0.398 & 0.498 \\
       \end{tabular}
       \caption{Results of emission coordinates of an event $\mathcal{R} = (t,x,y,z) = (0.1,0,0,0)$ 
in Eddington-Finkelstein coordinates in Schwarzschild spacetime.
The pulsars are freely falling toward a static black hole located at $(x_0, y_0, z_0) = (2.5, 0, 0)$.
$t$ denotes the Eddington-Finkelstein times when the pulsar world lines intersect with the observer's 
past light cone; at $t = 0$, all pulsars have zero velocity. $\tau$ lists the proper times: the emission 
coordinates of the reception event $\mathcal{R}$ can be recorded as the collection of any four of 
the $\tau^\alpha$. $x, y$, and $z$ are the Eddington-Finkelstein
spatial coordinates of the pulsars when the intersections occur.}
       \label{table:schwarzschild}
\end{table}

The numerical reinitialization described in Eq.(\ref{eq:numReinit}) was necessary for the 
stability of this curved space simulation. The eikonal
evolution suffered abrupt failures after $\sim 200$ integration steps; reinitializing 
every $75$ integration steps controlled this behavior.
Results were independent of the frequency of reinitialization, so long as it was no 
less frequent than every $75$ steps. 
%%%%%%%%%%%%%%%%%%%%%%%%%%

\section{Astrophysical Application: Fixing Gauge}
\label{sec:gauge}

Any set of four proper emission times may be used as emission coordinates, which we generically call $\zeta^\alpha$.
Since here we collect pulses from five pulsars, there are five such sets and we introduce a numbering 
${}_i\zeta^\alpha$ over all coordinate systems created by combining selected
$\tau^\beta$.
A typical set in our case might be: ${}_2\zeta^\alpha = \{ \tau^2, \tau^3, \tau^4,\tau^5\}$;
we restrict the possible choices by requiring the numbering to be ordered, increasing, in the quadruple. 
We note again that \cite{2009arXiv0905.3798T} shows how to combine results from 
more than four pulsar sources in a way that produces coordinates with reduced uncertainties. 

Since we are dealing with a particular event $\mathcal{R}$ in a given spacetime, all 
the ${}_i\zeta^\alpha$ are coordinate transformations of one another. Since the source 
pulsars are independent, the transformations are discontinuous: no continuous path of small 
transformations joins them. But besides these transformations, once a set of source pulsars is 
decided on, there is another group of continuous gauge transformations, which we will discuss below.

The emission coordinates---say $(\tau^1,\tau^2,\tau^3,\tau^4)$---form a quadruple of proper 
emission times of the measured pulses.
For now we assume that the pulses are closely enough spaced such that each $\tau^{\alpha}$ 
can be viewed as a continuous time signal determined to arbitrary precision.
(Our simulations made this assumption and we used other means---not pulse counting---to 
determine the source proper times.)
In practice one would interpolate; the interpolation process will be discussed briefly below.

It is clear that this coordinate system has a gauge group: affine transformations on each of 
the $\tau^\alpha \rightarrow A^\alpha \tau^\alpha +B^\alpha$ (no sum on $\alpha$), where 
$A^\alpha$ and $B^\alpha$ are finite and we restrict $A^\alpha$ to be positive.
Clearly $B^\alpha$ is an offset (e.g. Eastern time vs Pacific time) and $A^\alpha$ 
is a clock rate factor, or could be viewed as a function of the time unit chosen (e.g. seconds vs hours).

To construct a consistent coordinate system, $A^\alpha$ and $B^\alpha$ must be chosen 
and held fixed for all reception events.
(In our simulations we set $A^\alpha=1$, and defined $B^\alpha$ by demanding 
that $\tau^\alpha$ = 0 when the pulsar coordinate time $t=0$.)

Setting the gauge is intertwined with other steps in establishing the emission coordinates.
For practical purposes, one may choose to construct a consistent coordinate system, using one 
or more central master stations, by proceeding as follows:
\begin{enumerate}
\item Accumulate data on potential source pulsars. Compare pulse arrival stability against the 
best atomic clock. This will require the removal of known detector motion in the Solar System, 
and further polynomial fitting of the pulse arrival times. Transfer time of arrival solutions 
to a fiducial point, such as the barycenter of the Solar System.
\item Using good atomic clocks which hold stability over many pulse periods, interpolate the 
intervals between source pulses.
\item Define $\tau^\beta =0$ to correspond to the arrival of a specific pulse from each 
source $\beta$ at the receiver. It is intended that this is done at some finite specific 
time, while the pulse stability is being observed.
\end{enumerate}

Steps 1 and 2 provide precise interpolation into the intervals between source pulses, 
and partly imply a gauge (defining $A^\alpha$) for the coordinate system.
Step 3 defines the origin of the emission coordinate system, including fixing the offset 
$B^\alpha$. As the receiving station $\mathcal{R}$ ages and moves, its emission coordinates 
will move in a smooth way.

%%%%%%%%%%%%%%%%%%%%%%%%%
\section{Computational  Accuracy}
\label{sec:compAccuracy}

To provide a proof of principle, we have concentrated on the 
eikonal formalism and its generality of application; our numerical accuracy 
in examples is only moderate. By contrast the analysis of \cite{2009arXiv0912.4418D} 
(based on a different approach) is carried out essentially to machine 
precision for the Schwarzschild case. A number of approaches are being 
implemented to improve our code accuracy. The current code employs second-order discretization. This is being improved stepwise to fourth and 
ultimately to eighth order discretization. The current code is a unigrid code, 
so the same numerical error limit applies at all points of the grid. The backward 
wavefront (backward null cone) curvature is greatest near the 
event $\mathcal{R}$, but is small for most of the evolution from sources to $\mathcal{R}$. We will implement multiresolution so that the regions where the null cone is most highly curved are 
well resolved. Even with this improvement, we still require setting data on a 
small ``sphere" just to the past of the event 
$\mathcal{R}$. Care will be taken that this data setting is consistently convergent 
with the rest of the computation. Additionally we are studying  the behavior of 
the eikonal differential equation, to optimize for accuracy its 
translation into computational terms.

%%%%%%%%%%%%%%%%%%%%%%%%%

\section{Conclusion}
\label{sec:conclusion}
We have presented a robust numerical method to measure the emission coordinates of an
event in any generic spacetime configuration.
Our method uses a computational evolution of the eikonal equation describing the backward light cone.

We applied our method in two different numerical simulations: one in Minkowski spacetime and one in Schwarzschild spacetime.
In both simulations, we found that our methods are reliable in measuring the emission coordinates.
Errors in the measurements can be attributed to the numerical accuracy of the simulations, mainly due to the interpolation routine and the resolution of the three dimensional finite difference mesh.
We anticipate that with a higher resolution, we will be able to reduce the errors in our calculation of the emission coordinates.

Although these numerical simulations are preliminary, the same method can be used for pertinent problems, such as measuring the emission coordinates of the Earth, and therefore the Earth's trajectory.

%%%%%%%%%%%%%%%%%%%%%%%%%
\section{Acknowledgments}
\label{sec:acknowledgment}
We acknowledge DARPA and its XTIM team members for their significant contributions and suggestions for this work, especially the past DARPA program manager Derek Tournear, and NASA for its current support of XTIM technology development. We thank the anonymous referee for very helpful comments on this work.

\bibliography{biblio}

\end{document}